\documentclass{WileyMSP-template}

\begin{document}

\pagestyle{fancy}
%\rhead{\includegraphics[width=2.5cm]{vch-logo.png}}

\title{Phonon interference in the array of carbon nanotubes}

\maketitle

% Author: Please give full first and last names for authors and include * after the name of all corresponding authors

\author{Smirnov V.V.}
%\author{Author Two}
%\author{Author Three*}

% Dedication

\dedication{}
%Optional dedication here. If no dedication is required, please leave blank}

% Affiliations: Please provide adacemic titles (Prof. or Dr.) for all authors where applicable, and include an institutional email address for all corresponding authors
\begin{affiliations}
V. V. Smirnov\\
%Address
Federal Research Center for Chemical Physics, RAS\\
4 Kosygin street, 119991, Moscow, Russia\\
Email Address: vvs@polymer.chph.ras.ru

%A. N. O. Author\\
%Address

\end{affiliations}

% Keywords: Please provide a minimum of three and a maximum of seven keywords, separated by commas


% Abstract should be written in the present tense and impersonal style (i.e., avoid we), and be at most 200 words long
\begin{abstract}

The dynamics of the one-dimensional array of the single-walled carbon nanotubes, which interact by van der Waals forces, is considered.
The molecular dynamics simulation shows that both the mutual displacements of the nanotubes and the deformations of their walls occur in the low-frequency oscillations domain.
The composite model taking into account both types of the nanotubes' motions was developed in the framework of the thin elastic shell theory.
Such an approach allows us to reduce the problem to the dynamics of the two-parametric linear lattice with contact interaction.
The dispersion relations are represented analytically and the multi-channel propagation that results to phonon interference (the acoustical analogue of the Fano resonance) is observed in the presence of the array's irregularities.
The latter can be formed with redundant nanotubes, which are placed over the array in the groove between neighbour sites.
The calculations of the transmittance have been performed by the transfer matrix method for several typical configurations.

\end{abstract}

% Text: Please use section headings and subheadings as specified below. For communications, all section headings apart from Experimental Section should be removed
% Please make the first reference to a display item bold: \textbf{Figure 1}
% Do not abbreviate Figure, Equation, etc.; display items are always singular, i.e., Figure 1 and 2.
% Equations are always singular, i.e., Equation 1 and 2, and should be inserted using the {equation} environment, not as graphics
% Please do not use footnotes in the text, additional information can be added to the Reference list.
\section{Introduction}
Carbon nanotubes (CNTs) are the one of most known carbon allotropes, which have been widely studied since their discovery at 1991 \cite{Iijima1991}.
Their unique mechanical, electrical, optical, and thermal properties give rise to the interest to them in the various fields of the biology, chemistry, electronics, and material science \cite{Rao2018,And02005,Dresselhaus1998}.
Their perspective applications for the development of the nanoscale electronic devices, biosensors, nanocomposite materials need in the knowledge of the properties of the isolated nanotubes as well as of the arrays of them \cite{Rao2018}.
%Moreover, the carbon nanotubes are often considered as the one-dimensional objects, which allows us to understand many fundamental problems (REF).
Many fields of the science and technology are in interest in the collective properties of the nanotubes when their interaction is essential.
It relates to the problem of the design of the new nanoscale composite materials, in particular, the materials, which contain the nanotube ropes and mats.
%The nanotube interaction may be important at the development of the nanoscale electronic devices with the nanotubes placed on the surface of the substrate \cite{Park2013}.
%The  interaction of nanotubes as well as their interaction with the substrate changes the frequency characteristics of the CNTs \cite{Henrard1999,Sauvajol2002}.
%It is known that the deformation of the nanotubes on the surface of the elastic substrate depends on the presence of another nanotubes \cite{Xiao2008}.
The development of the methods for the fabrication of the aligned nanotubes also gives rise to the interest to the nanotube interactions \cite{Zhang2017,Xiao2009}.
The interaction between nanotubes as well as their interaction with the substrate has the effect on the nanotubes' deformations, frequency characteristics, electronic band structure and charge carrier mobility \cite{ Henrard1999,Sauvajol2002,Xiao2008,vanderGeest2011,Perebeinos2009,Flebus2020}.
These effects are important for the development of the electronic and electro-mechaniccal devices with usage of the nanotubes \cite{Park2013}.

The interaction of the CNTs placed at some distance from one to other results from the van der Waals forces acting between carbon atoms, which form the nanotubes \cite{Girifalco2000,Siber2009}.
In spite of the van der Waals interaction is not strong, its effective radius is large enough that leads to the collective character of the nanotube interaction, when many carbon atoms contribute to the energy.
In order to avoid the infinite summation the effective cut off distance has been usually introduced in the procedure of the numerical simulation \cite{Harik2018,Savin2020}.
The alternative approach consists in the continualization of atom distribution over the CNT's surface with the density, which is determined by the area of the elementary cell 
%($\rho_s= 4/3\sqrt{3} a^2; a=1.42 A$)
 \cite{Harik2018,RafiiTabar2008}.
Such an approach allows us to take into account the mutual arrangement of the CNTs but the final values of the interaction energy can be obtained by the numerical estimation the respective integrals only.
This procedure has been used in the number of works \cite{Sun2005,Sun2006,Popescu2008,Zhao2013} starting from the paper by Girifalco \cite{Girifalco2000}, where the concept of the "universal carbon" potential has been made.
One should notice that the researches mentioned deal with the interaction of the nanotubes without any deformations.
However, the observation of the CNTs in the bundles shows that some "polygonization", i.e. the difference the nanotubes` cross section from the ideal circle, occurs \cite{Tang2000}.
The packaging of the CNTs in the bundles corresponds to the dense hexagonal arrangement, the symmetry of which controls the mode of the nanotube deformation.
If the nanotubes are long enough the bundle can be considered as a fragment of two-dimensional regular array, called the CNT-crystal, which has been studied in work \cite{Tersoff1994} for the first time.
The elastic properties of the CNT crystal were considered in \cite{Popov2000, Saether2003a,Saether2003b}.
It is important that the total energy of the CNT crystal consists of the energy of nanotube interaction as well as the energy of their deformation \cite{Smirnov2019}.
The nanotube deformation in the CNT crystal should be considered as the internal degree of freedom, the presence of which gives rise to the optical-type branch in the dispersion relation.
The description of the CNT crystal dynamics has to include the mutual displacement of the nanotubes' center of masses as well as their deformation oscillations.
The latter may be studied in the framework of the elastic thin shell theory \cite{Amabili2008B}, when the CNT are considered as a thin elastic shell, which is characterized by the elastic moduli, Poisson ratio and the effective thickness of the "wall" \cite{Vodenitcharova2003,Huang06,Gupta2010,Chang10}.
One should notice that such an approach to study of the nanotube vibrations is often used in the problem of the bending vibrations with or without elastic foundation \cite{CYWang04,Liew07}.
Moreover, it was shown that there is well agreement between the data of the molecular-dynamical simulations and the results of the description of the nanotubes in the framework of the nonlinear theory of thin elastic shell by Sanders and Koiter \cite{Silvestre11,Silvestre12,Rafiee2014}.
The latter turn out to be successful in the analysis of the low-frequency oscillation localization in the single-walled CNT \cite{Smirnov2014,Smirnov2016PhysD} as well as of the interaction of nonlinear normal modes which belong to the different branches of the dispersion relation \cite{Smirnov2018}.
Generally speaking, the problem of the thin shell deformation in the nonlinear formulation is one of most difficult one in the contemporary mechanics and it may be solved analytically in isolated cases only \cite{Amabili2008B}.
However, the mentioned deformation of the CNTs is specific in that the changing the nanotubes cross section, which is normal to the tube`s axis, is not accompanied the variation of its contour length.
The latter leads to some relationship between radial and circumferential displacements of the shell that allows us to reduce the complexity of the dynamical  problem \cite{Kaplunov2016,Smirnov2016PhysD}.
At a small deformation of the nanotubes the changing cross section`s contour is characterized by the circumferential wave number $l \geq 0$:
\begin{equation}\label{eq:contour}
R(\theta) = R_0 \left(1+\sum_l{w_l \cos{l \theta}} \right),
\end{equation} 
where $R_0$ is the radius of non-deformed nanotube, $\theta$ is the azimuthal angle and $w_l$ is the amplitude of the radial displacement of the $l$-th mode.
For the isolated nanotube the circumferential wave numbers $l=0$ and $l=1$ correspond to the well known radial breathing mode (RBM) and bending oscillations, respectively, while $l=2$ gives rise so-called circumferential flexure mode (CFM), which is the most low-frequency optical-type vibration of nanotubes \cite{Dresselhaus00,Mahan02}.
%The reason of the  low frequency is that such vibrations are coupled to the deformation of the conformation angles and are not associated with variation of the valence bonds or the valence angles.
Further we consider the particular system of the one-dimensional array of the single-walled nanotubes that turns to be interesting as the model one and may be useful in various problems of the nanoelectronics and nanomechanics.

\section{The model}
\begin{figure}
 a) \includegraphics[width=\linewidth]{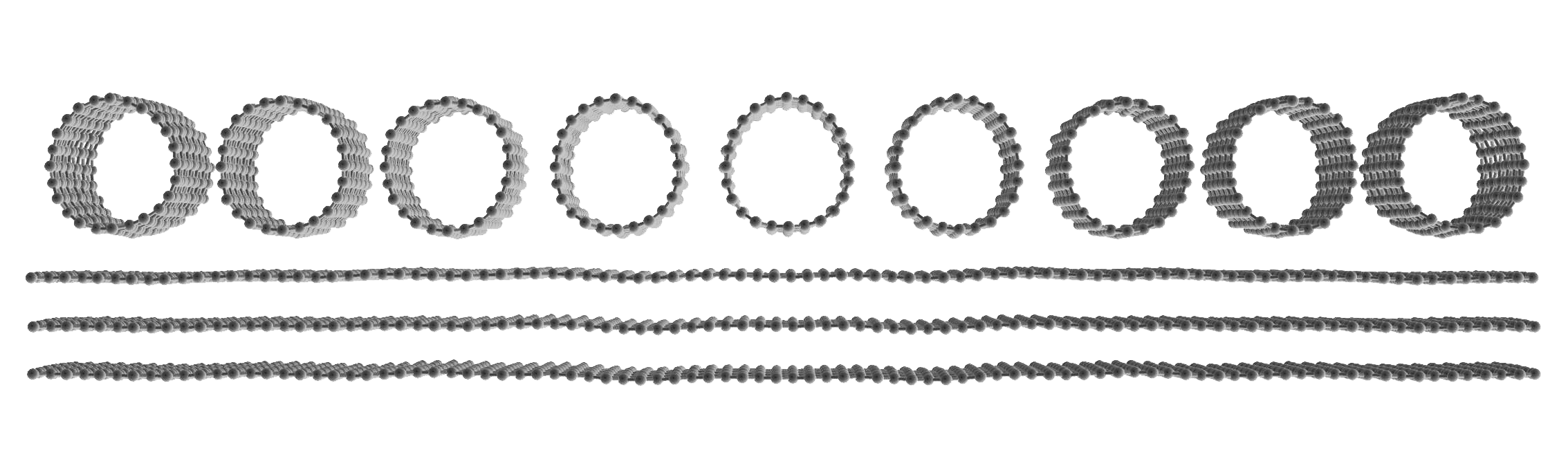}\\
  b)\includegraphics[width=\linewidth]{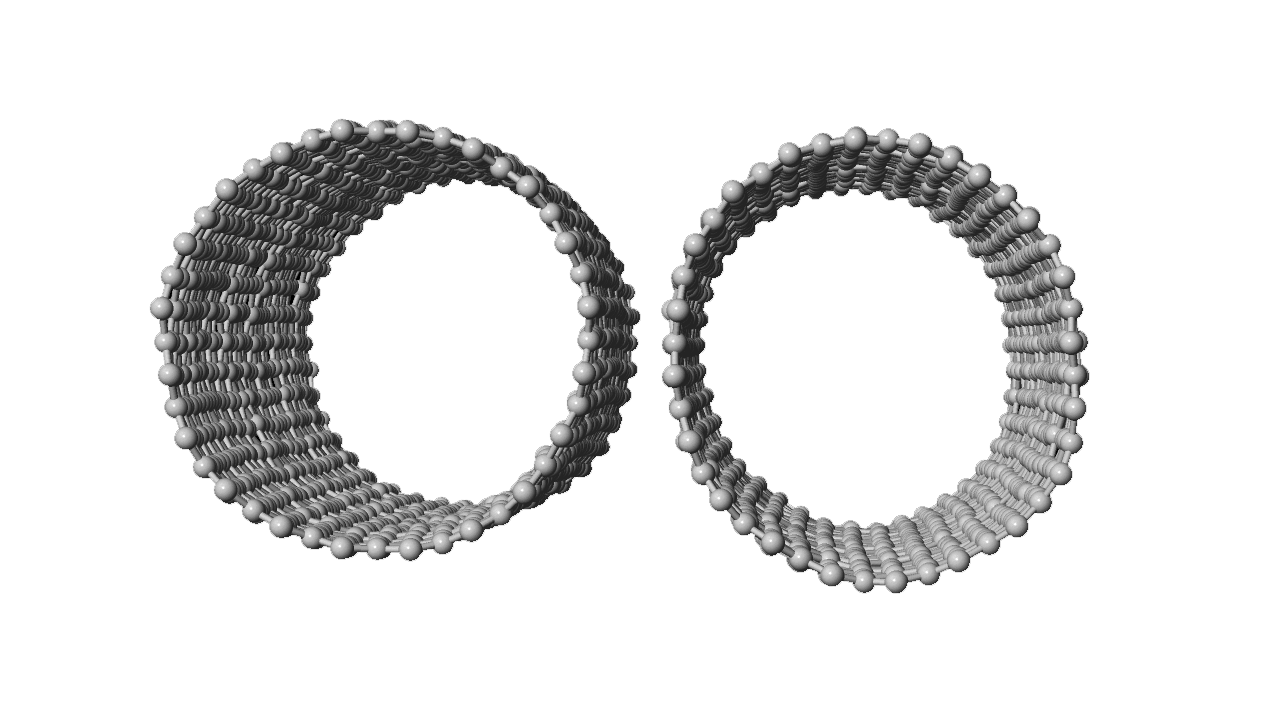}
  \caption{a) The snapshot of MD simulation of the (12,0) CNT  array on the tree-layered graphene. b) The snapshot of MD simulation of two (20,0) CNTs. }
  \label{fig:MDsim1}
\end{figure}

\begin{figure}
  \includegraphics[width=\linewidth]{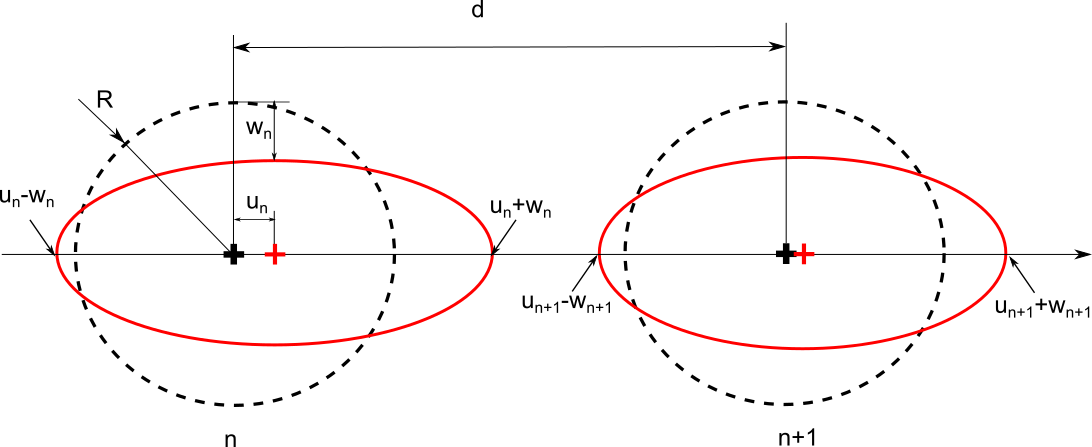}
  \caption{Sketch of two nanotubes interaction. Dashed contours correspond to non-deformed CNTs of radii $R$ at equilibrium distance $d$ in the regular array. Red contours show the deformed nanotubes and the displacements of their center of masses (red crosses). Variables $u$ and $w$ correspond to the displacement of center of masses and amplitude of the radial deformation, respectively.}
  \label{fig:f1}
\end{figure}

\begin{figure}
  \includegraphics[width=\linewidth]{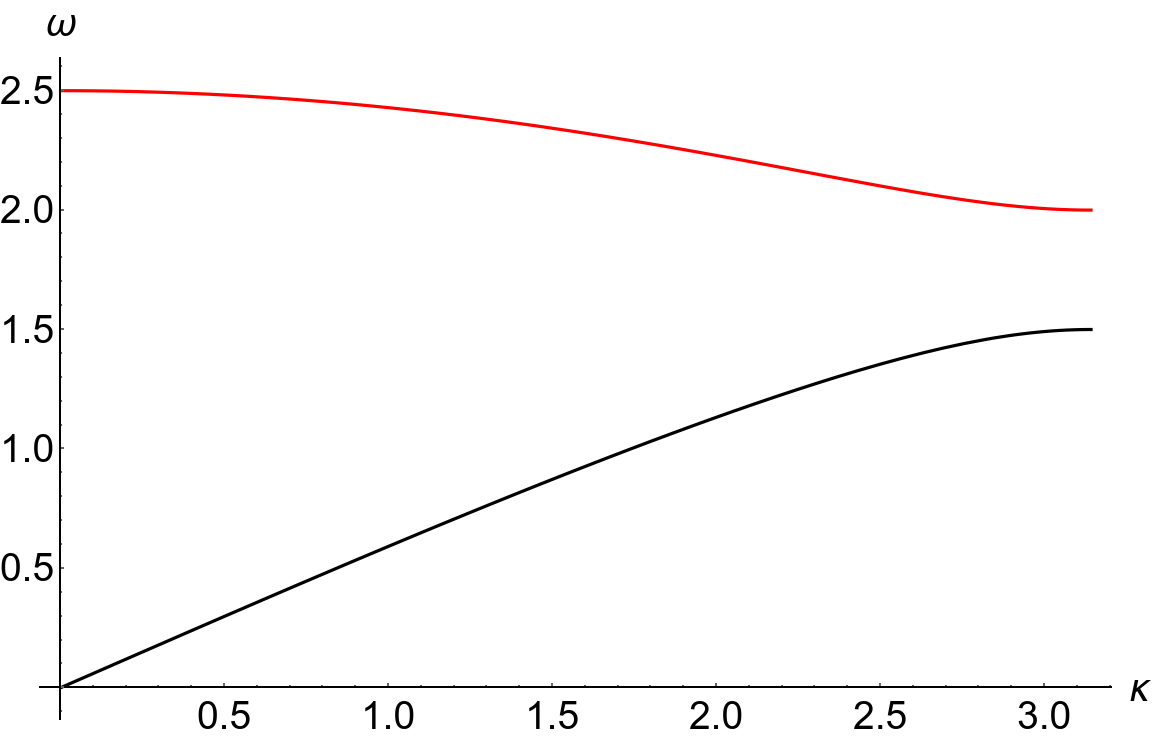}
  \caption{Dispersion curves for equations (\ref{eq:eqm0}). $\chi=1, \, \Omega=1.5$ }
  \label{fig:DR}
\end{figure}

Let us consider the one-dimensional array of the single-walled CNTs, which are placed on some distance $d$ from each  other.
In the equilibrium, the nanotubes' interaction results to that the cross section's contour can undergo the deformations \cite{Smirnov2019} which is described by Equation (\ref{eq:contour}) with some set of the circumferential wave numbers $l$.
Figure \ref{fig:MDsim1}(a) shows the snapshot of the molecular dynamics simulation of the (12,0) CNT array of the surface on the three-layered graphene at the temperature $T=300 K$.
The circumferential flexure deformations of the CNTs' walls result to the imperfection of the nanotube cross sections.
Figure \ref{fig:MDsim1}(b) shows the snapshot of the simulation of the interaction two (20,0) nanotubes at  $T=300 K$.
The quasi-elliptical deformation of the right-hand nanotube is well observed.
From the viewpoint of the nanotube interaction the energy of the system consists of the energy of elastic deformation of the CNT and the energy of the van der Waals interaction between carbon atoms belonging to neighbouring nanotubes.
The first is determined by the CNT's circumferential rigidity and may be describes as the on-site described as follows:
\begin{eqnarray}
E_c=\frac{\Omega^2}{2} w^2,
\end{eqnarray}
$w$ is the amplitude of the radial deformation and $\Omega$ is the frequency of natural oscillations of the nanotubes which accompanied by varying of the nanotube's cross section (see Supporting Information).
In the one-dimensional array the symmetry dictates the circumferential flexure mode with $l=2$ as the preferential one.
We assume that the energy  of the van der Waals interaction between neighbour nanotubes is determined by the distance between nanotubes' walls, which depends on the displacement of center of masses as well as on the radial deformation amplitude.
Figure \ref{fig:f1} shows the sketch of the interaction of two deformed CNTs.
It is essential that the effect of the circumferential deformation on the inter-wall gap  differs for the left and right "edges" of the nanotubes.
On the right-hand edge of the nanotube the radial deformation and displacement of center of masses are summarized, while on left-had edge they effect in the opposite direction.
Under these assumptions we can represent the potential energy of the regular array of the nanotubes in the linear approximation as follows:
\begin{eqnarray}\label{eq:Vpot1}
V=\frac{1}{2} \chi ^2 \left(\left(\left(u_n-u_{n-1}\right)-\left(w_{n-1}+w_n\right)\right){}^2+\left(\left(u_{n+1}-u_n\right)-\left(w_{n+1}+w_n\right)\right){}^2\right)+\frac{\Omega ^2}{2} w_n^2,
\end{eqnarray}
where constant $\chi$ characterizes the rigidity of the van der Waals interaction, and  frequency $\Omega$ is determined by the own rigidity of the nanotube's contour.
Here we use the dimensionless variables: displacement $u$ and radial deformation $w$ are measured in the units of the nanotube's radius, frequency $\Omega$ and coupling constant $\chi$ are related with the frequency of the radial breathig mode (RBM). 
(One should note, that frequency $\Omega$ can take into account the effect of the substrate attraction, if the array is placed on the solid surface \cite{Henrard1999}.)
Let us note that displacement of the center of masses $u_n$ and amplitude of radial displacement $w_n$ of the n-th nanotube are represented by different manner in the first and second terms of Equation \ref{eq:Vpot1}, as it was  mentioned above.
Therefore, we define new variables, which describe the displacements of the left and right "edges" of the nanotube as follows:
\begin{eqnarray}\label{eq:edges}
\psi_n=\frac{1}{\sqrt{2}} \left( u_n - w_n \right), \quad \varphi_n =\frac{1}{\sqrt{2}} \left( u_n + w_n \right),
\end{eqnarray}
where factor $\sqrt{2}$ is introduced for the convenience.
So, in terms of these variables the energy of the system can be written in the form
\begin{eqnarray}\label{eq:E0}
E=\sum_n{\frac{1}{2} \left( \left(\frac{d \psi_n}{d t}\right)^2+\left(\frac{d \varphi_n}{d t} \right)^2 \right)+\chi^2 \left( \left(\psi_{n+1}-\varphi_n \right)^2+ \left( \psi_n-\varphi_{n-1} \right)^2 \right) +\frac{\Omega^2}{4} \left( \varphi_n-\psi_n \right)^2}.
\end{eqnarray}
The respective equations of motion are read as
\begin{eqnarray}\label{eq:eqm0}
\frac{d^2 \varphi_n}{d t^2}+\frac{\Omega^2}{2}\left(\varphi_n-\psi_n \right)+2\chi^2 \left(\varphi_n-\psi_{n+1}\right) =0 \\ \nonumber
\frac{d^2 \psi_n}{d t^2}+\frac{\Omega^2}{2} \left(\psi_n-\varphi_n \right)+ 2 \chi^2 \left(\psi_n-\varphi_{n-1} \right) =0.
\end{eqnarray}
The dispersion relations consist of two branches:
\begin{eqnarray}\label{eq:DR0}
\omega^2=\frac{1}{2}\left( 4 \chi ^2+\Omega ^2\pm \sqrt{8 \chi ^2 \Omega ^2 \cos (\kappa )+16 \chi ^4+\Omega ^4} \right)
\end{eqnarray}
Figure \ref{fig:DR} shows  dispersion relations (\ref{eq:DR0}) for the CNT array with parameters: $\chi=1/0, \Omega = 1.5$.
One should remark that the right edge of the optical branch of dispersion relation (\ref{eq:DR0}) has to correspond to the frequency  of the natural oscillations of the isolated nanotube $\Omega$, while the acoustic branch has to converge to value $2\chi$.
However, it occurs if frequency $\Omega$ is smaller than $2\chi$.
Otherwise, we can observe $\omega \rightarrow 2\chi$ for the optical branch and $\omega \rightarrow \Omega$ for the acoustical one (see Figure \ref{fig:DR}).

\begin{figure}
  \includegraphics[width=\linewidth]{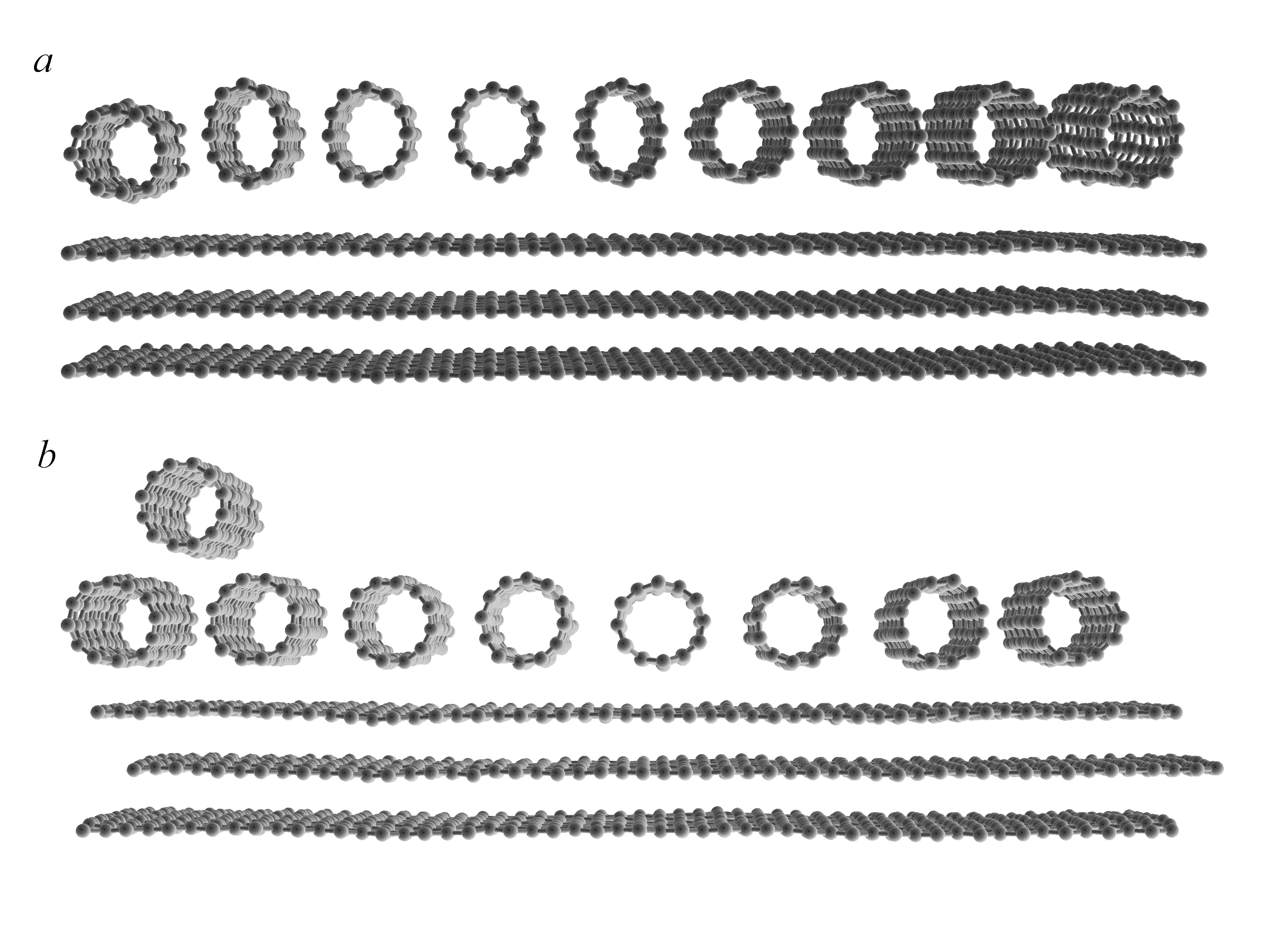}
  \caption{ The snapshots of the MD simulation of the (12,0) CNTs on three-layered graphene under external stress along the graphene surface and normally to the nanotubes' axes. Panels (a) and (b) show the configuration before and after loss of the stability.}
  \label{fig:defectarray}
\end{figure}

\begin{figure}
  \includegraphics[width=\linewidth]{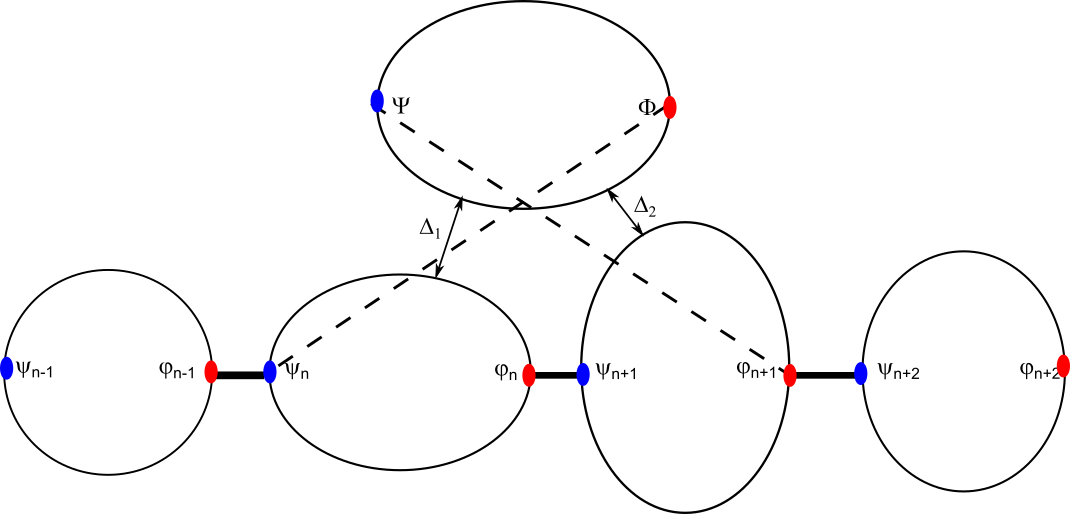}
  \caption{Sketch of the CNT array with additional nanotube as the "discrete state". Thick solid lines show the "contact" interaction and dashed lines show the bonds between regular array and "discrete state" nanotube.}
  \label{fig:DS01}
\end{figure}

In the CNT array the phonon interference arises as  the result of the  Fano resonance \cite{Kosevich2008,Miroshnochenko2010}, if an additional nanotube is placed in the groove between two neighbour nanotubes of the array.
Such a "discrete state" can be formed artificially or be the result of instability of the array under action of the pressure in the direction, which is normal to the nanotubes' axes.
The example of such instability is shown in Figure \ref{fig:defectarray}.
The "excess" nanotube in Figure \ref{fig:defectarray}b arises as the result of the instability of the  initially stressed array in Figure \ref{fig:defectarray}a.
One of the nanotubes is ejected from the array under action of the thermal fluctuations and sites the position in the groove between two neighbour nanotubes.
Such a configuration turns out to be stable and the upper nanotube does not change its location.
The sketch of intertubes' bonds in the fragment of the CNT array with the "discrete state" is represented in Figure \ref{fig:DS01}.
The "discrete state" and the nanotubes of the array interact by the van der Waals forces and the energy of this interaction is controlled by distances $\Delta_1$ and $\Delta_2$. 
One can show that the values of these distances depend on the differences $(\Psi-\varphi_{n+1})$ and $(\Phi-\psi_n)$ and the energy of the "discrete state" can be written in the form: 
\begin{eqnarray}\label{eq:DSenergy1}
V_d=\frac{\chi ^2}{4} \left(\left(\Psi -\varphi _{n+1}\right){}^2+\left(\Phi -\psi _n\right){}^2\right)+\frac{\Omega_1 ^2}{4} (\Phi -\Psi )^2
\end{eqnarray}
The "discrete state" has two own frequencies:
\begin{eqnarray}\label{eq:DSfrequency1}
\omega_1 = \frac{\chi}{\sqrt{2}}, \quad \omega_2 = \sqrt{\Omega_1 ^2+\frac{\chi ^2}{2}}
\end{eqnarray}
   
%\subsection{First Subsection}
In order to study the transmission of the wave through the "discrete state" placed between sites $n$ and $n+1$ we use the transfer matrix method \cite{Tong1999}.
Assuming that variables $\psi$ and $\varphi$ depend on the time as $~ e^{i \omega t}$ we can represent Equations (\ref{eq:eqm0}) in the vector form: 
\begin{eqnarray}\label{eq:regarray1}
\left(
\begin{array}{c}
\psi_{n+1}\\
\varphi_n
\end{array}
\right)
=T_0 \left(
\begin{array}{c}
\psi_n \\
\varphi_{n-1}
\end{array}
\right)
\end{eqnarray}
where transfer matrix $T_0$ is read as follows (see Supporting Information):
\begin{eqnarray}\label{eq:TMregular}
T_0=\left(
\begin{array}{cc}
 \frac{\left(2 \chi ^2-\omega ^2\right) \left(2 \chi ^2-\omega ^2+\Omega ^2\right)}{\chi ^2 \Omega ^2} & -\frac{4 \chi ^4-2 \chi ^2 \omega ^2+\chi ^2 \Omega ^2}{\chi ^2 \Omega ^2} \\
 -\frac{-4 \chi ^2+2 \omega ^2-\Omega ^2}{\Omega ^2} & -\frac{4 \chi ^2}{\Omega ^2} \\
\end{array}
\right)
\end{eqnarray}
The coupling between sites $n+1$ and $n-m$ is described by the relation:
\begin{eqnarray}\label{eq:regarray2}
\left(
\begin{array}{c}
\psi_{n+1}\\
\varphi_n
\end{array}
\right)
=Z \left(
\begin{array}{c}
\psi_{n-m+1} \\
\varphi_{n-m-1}
\end{array}
\right),
\end{eqnarray}
where $Z$ is the transfer matrix of the $m-$steps way and $Z=T_0^m$ for the regular (defectless) array.

However, if the array's fragment contains the irregularities, the one-step matrix $T_{i,j}$  differs from matrix  $T_0$.
Thus, we should calculate the transfer matrix taking into account the interactions between nanotubes of the regular array and the redundant nanotube, which forms the "discrete state".
In particular, the transition from $\left(\begin{array}{c}\psi_{n+3}\\\varphi_{n+2}\end{array} \right)$ to $\left(\begin{array}{c}\psi_{n-1}\\\varphi_{n-2}\end{array} \right)$ is described by the matrix
\begin{eqnarray}\label{eq:Zmatrix0}
Z=\left(I-T_0 T_{2,1} \tau_{1,2} T_0^{-1} \right)^{-1} T_0 \left(T_{2,1} T_{1,0}+\tau_{2,0} \right)T_0,
\end{eqnarray}
where $I$ is identity matrix and the  transfer matrixes $T_{i,j}$ and $\tau_{i,j}$ describe the transition from the bond $\{n+i,n+i-1\}$ to $\{n+j,n+j-1\}$ (see Supporting Information for details).
They are calculated with accounting the oscillations of the "discrete state" nanotube.
%Because of their representation is extremely bulky, the respective calculations are put in the Supporting Information.

In order to analyse the transmission of the wave in the system under consideration, we assume that the left half-array contains both the incoming and reflected wave, while only the transmitted wave occurs in the right half-array.
Thus, we should write
\begin{eqnarray}\label{eq:waves}
\left(\begin{array}{c}\psi_{n-j}\\\varphi_{n-j}\end{array} \right) =A_0 e^{i \kappa \left(n-j\right)}+A_r e^{-i \kappa \left(n-j \right)}, \quad j > 1  \\ \nonumber
\left(\begin{array}{c}\psi_{n+j}\\\varphi_{n+j}\end{array} \right)=  A_t e^{i \kappa \left( n+j \right)}, \quad\quad\quad\quad\quad\quad\quad j>3,
\end{eqnarray} 
where $A_0$ and $A_r$ are the two-component vectors of the amplitudes of the incoming and reflected waves, and $A_t$ is the vector of amplitude of the transmitted wave.
Amplitudes $A_0, A_r$ and $A_t$ are the vectors because the array under study is  complex  and the solution for  Equations (\ref{eq:eqm0}) contains two component.
It should be taken into account in the process of finding the transmission coefficient, which can be defined as the square of the modulus of the ratio of amplitudes of the incoming and transmitted waves.
Under these conditions the transmission coefficient $t=|A_t/A_0|^2$ can be written as follows:
\begin{eqnarray}\label{eq:transmission}
t=\frac{4 \left(\Omega ^4+8 \chi ^2 \Omega ^2 \cos{\kappa}+16 \chi ^4\right) \Omega ^4 \sin ^2{\kappa}}{| \left(\Omega ^2+4 e^{i \kappa } \chi ^2\right) \left(4 \chi ^2-2 \omega ^2+\Omega ^2\right)\left(Z_{12}-Z_{21}\right)-e^{i \kappa }  \left(4 \chi ^2-2 \omega ^2+\Omega ^2\right)^2 Z_{22}+e^{-i \kappa }  \left(\Omega ^2+4 e^{i \kappa } \chi ^2 \right)^2 Z_{11} |^2 },
\end{eqnarray}
where $Z_ij$ are the components of the matrix $Z$.
One should note that transmission coefficient (\ref{eq:transmission}) is similar to that in \cite{Tong1999}, but is not  identical with it because of the complexity of the CNT array.

\section{Phonon interference}
The model discussed above has two parameters: the constant of the intertube interaction $\chi$ and frequency of natural oscillations $\Omega$.
The theoretical estimation of constant $\chi$ can be made by the direct numerical evaluation of the energy of van der Waals interaction \cite{Sun2005,Sun2006,Smirnov2019}.
The parameters of the Lennard-Jones potential are well known for the carbon nanostructures \cite{Girifalco2000}.
The frequency of the circumferential flexure oscillations $\Omega$ can be evaluated in the framework of the thin elastic shell theory \cite{Mahan02,Chico06,Kaplunov2016,Smirnov2016PhysD} with using of the effective elastic constants of the nanotubes or can be obtained from the data of the molecular-dynamical simulations.
In particular, the dispersion curves for the nanotube array in work \cite{Savin2021} can be well approximated with ratio $\Omega/\chi \approx 1.55$ and dimensional value of the frequency $\Omega_d \approx \,80 cm^{-1}$.
The measured value of the frequency of the circumferential flexure mode is $27 \, cm^{-1}$ for the separated (10,0) nanotube \cite{Dresselhaus00} and $\sim 80 cm^{-1}$ for the nanotube in the bundle \cite{Sauvajol2002}.
Further we will use the dimensionless value $\chi=1$ and the ratio $\Omega/\chi = 1.5$ for the calculation of the transmission coefficient.

\begin{figure}
  \includegraphics[width=\linewidth]{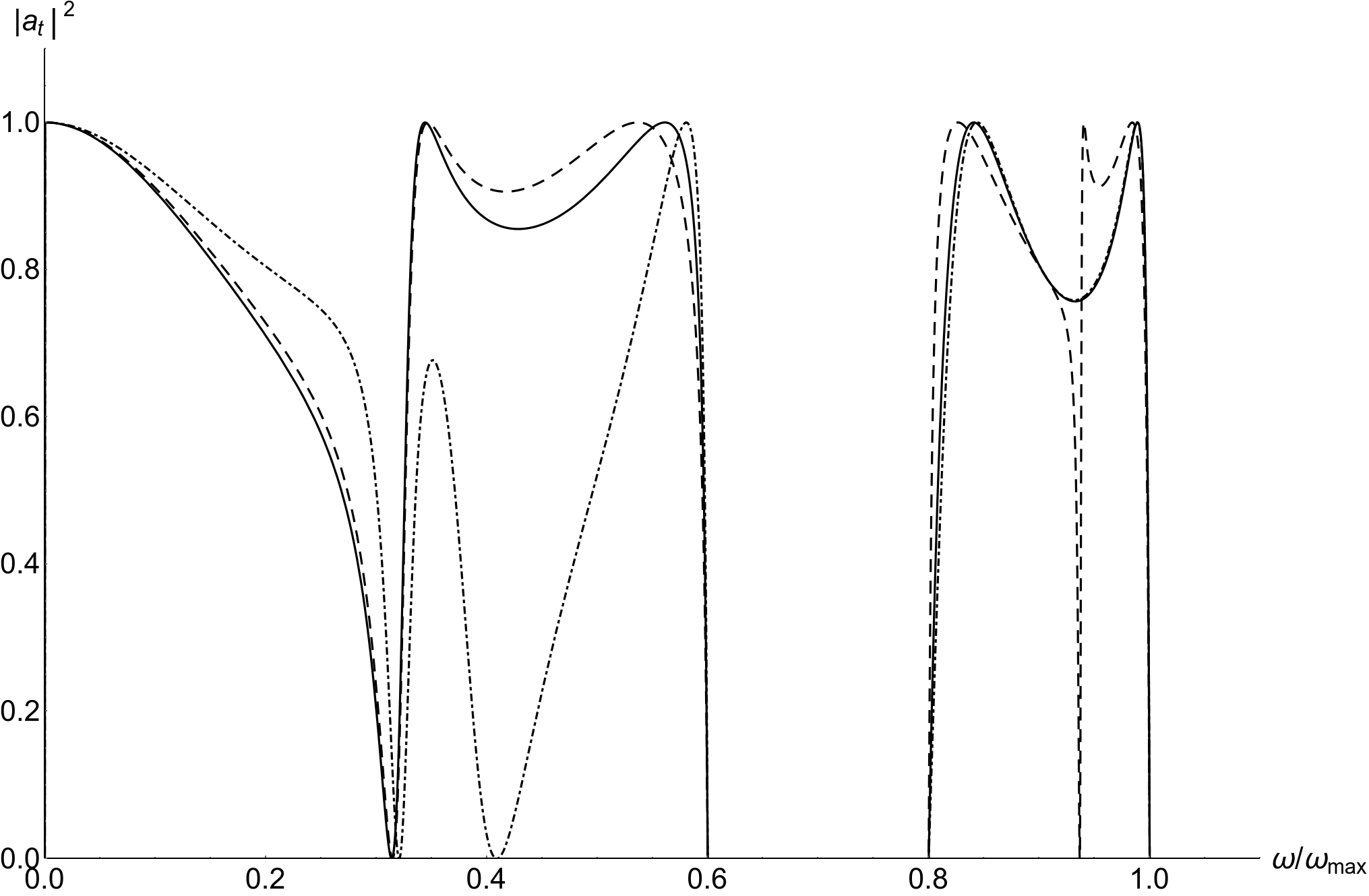}
  \caption{Normalized amplitude of transmitted wave vs normalized frequency for single nanotube on the regular array. Solid, dashed and dot-dashed curves correspond to $\Omega_1/\Omega = 1, 1.5$ and $0.5$, respectively.
  % Red dashed line shows the low frequency resonance of discrete state; blue, green and magenta lines represent the high-frequency resonances of discrete state with $\Omega_1/\Omega = 0.5, 1.0$ and $1.5$, respectively. 
  }
  \label{fig:single}
\end{figure}

As it was mentioned above, the simplest structure, which leads to the Fano resonance in the CNT array, is the single "redundant" nanotube ejected from the regular array (see Figure \ref{fig:defectarray}.b).
However, a similar structure with the redundant nanotube, which parameters differ from the parameters of the array's CNTs, can leads to the phonon interference also.
The nanotubes of smaller diameter has the bigger rigidity and the larger frequency $\Omega_1 > \Omega$.
And vice versa, a larger nanotube has a smaller frequency of the circumferential flexure oscillations.
If these frequency are in the permitted band we should observe the effect of the phonon interference.
Figure \ref{fig:single} shows the transmission coefficients for three structures with different redundant nanotube as the function of the reduced frequency $\omega/\omega_{max}$, where $\omega_{max}=\sqrt{4 \chi ^2+\Omega ^2}$ is the frequency of the optical branch at wave number $\kappa=0$. 
Solid, dashed and dot-dashed curves correspond to the redundant nanotubes with natural frequencies $\Omega_1
= \Omega, 1.5 \, \Omega$ and $0.5 \, \Omega$, respectively.
One can see that all three structures have the destructive interference that leads to full reflection of the incoming phonons at the relative frequency $\omega/\omega_{nax} \approx 0.3$.
This frequency belongs to the acoustic branch of the spectrum and corresponds to oscillations of the redundant nanotube as whole.
Second resonant frequency, which associates with circumferential flexure oscillations of the excess nanotube, is in the forbidden band if $\Omega_1=\Omega$.
If the redundant nanotube is more rigid than the CNTS of the array, its resonant frequencies are in the acoustical as well as in the optical branches (see Figure \ref{fig:single} - dashed curves).
Thus, such a nanotube reflects both the acoustical and optical phonons.
While the resonance in the acoustical part of the spectrum is similar to one for the nanotube with $\Omega_1=\Omega$, the destructive resonance in the optical branch is essentially more sharp.
If the nanotube above the array is larger than the CNT of the array, its natural frequency is less than $ \Omega $. 
The dot-dashed line in Figure \ref{fig:single} shows the transmittance for a nanotube with an eigenfrequency $ \Omega_1 = 0.5 \, \Omega $. 
In this case, both resonant frequencies are in the acoustic region, and we can observe a certain overlap of the destructive resonances.
 
\begin{figure}
  \includegraphics[width=\linewidth]{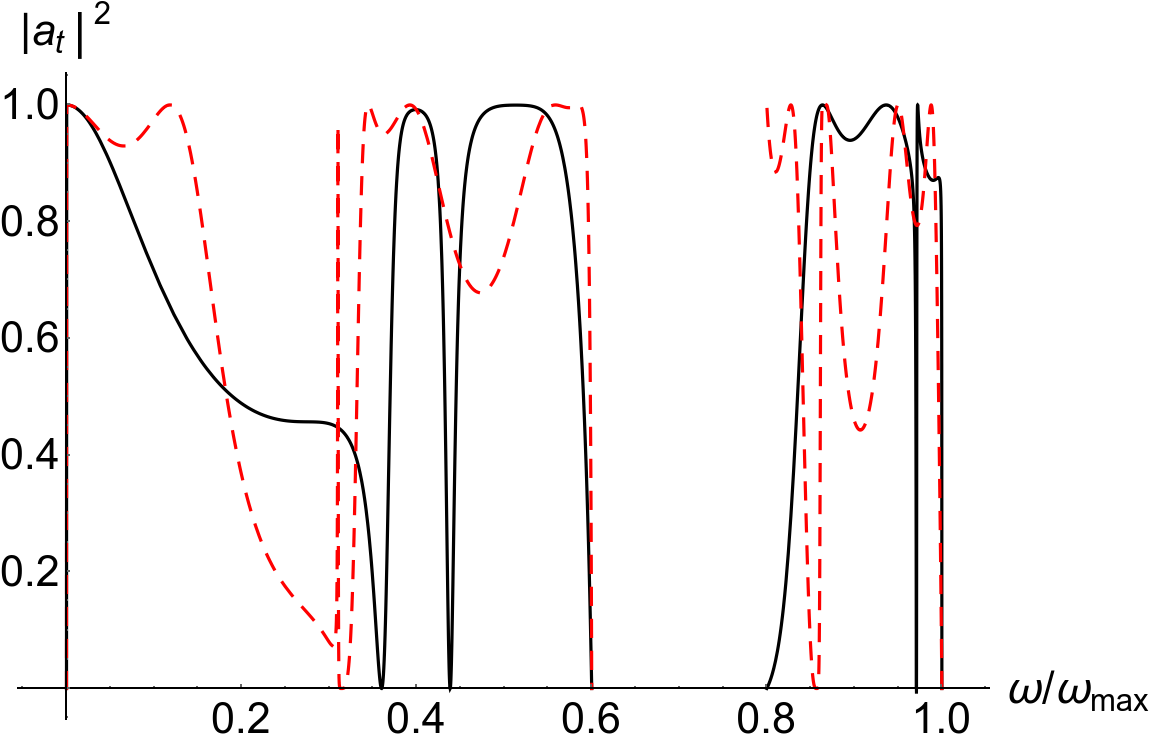}
  \caption{Normalized amplitude of transmitted wave vs normalized frequency for various configuration of the nanotubes on the regular array. Solid black and dashed red  curves curves correspond to doubled and two separated nanotubes on the regular array, respectively. Parameters: $\chi=1.0$, $\Omega=1.5$, $\Omega_1=1.5$.
  % Red dashed line shows the low frequency resonance of discrete state; blue, green and magenta lines represent the high-frequency resonances of discrete state with $\Omega_1/\Omega = 0.5, 1.0$ and $1.5$, respectively. 
  }
  \label{fig:double}
\end{figure}

Another structure that can lead to phonon interference in the CNT array is a combination of several nanotubes above the array.
For example, two additional nanotubes can be located in adjacent grooves (double CNTs) formed by three consecutive nanotubes of the array, or they can be placed at some distance from each other (separate CNTs).
In the first case, the connections between the additional CNTs and the nanotubes of the array overlap, and the resulting transfer matrix $Z$ has a more complex structure.
(Some details of these configurations are presented in the Supporting Information.)
The second combination should be considered as two non-interacting resonant structures separated by a fragment of a regular lattice.
In this case, we can construct the resulting transfer matrix as a combination of the matrix $Z$ of the Equation (\ref{eq:Zmatrix0}) and the product of matrices $T_0$.
Figure \ref{fig:double} shows the examples of the transmittance for resonant structures with two nanotubes.
The solid line corresponds to the doubled nanotubes above the array, while the dashed curve is associated with two nanotubes, which are located on three lattice constants from one to the other.
We can observe the effect of both destructive and constructive resonance in both the acoustic and optical regions.
The main acoustic destructive resonance near the frequency $ \omega \approx 0.3 \omega_{max}$ always occurs, but in the case of separated nanotubes, an extremely narrow constructive resonance arises in the vicinity of it.
The transmittances in Figure \ref{fig:double} have been calculated for the redundant nanotubes, which are the same as CNTs in the array.
Nevertheless, the resonances in the optical domain appear as for doubled as well as for the separated nanotubes, therefore the optical phonons of  certain frequencies are reflected from the considered structures.
Thus we can effectively control the phonon transmittance through the CNT array by the various combination of the additional nanotubes placed over the array.

%\begin{eqnarray}
%Z=\biggl( \left(T_0 (T_{32}T_{21}+\tau_{31})\tau_{12}(I-T_{21}\tau_{12})^{-1}(T_{21}T_{10}+\tau_{20})+\\ \nonumber
%T_0 (T_{32} T_{21}+\tau_{31})T_{10}+T_0 (\tau_{30}+T_{32}\tau_{20})\right)T_0 \biggr)^{-1} \multiply \\ \nonumber
%\biggl( I-T_0(T_{32}\tau_{23}+(\tau_{31}+T_{32}T_{21} )\tau_{13})T_0^{-1}- \\ \nonumber
%T_0(\tau_{31}+T_{32}T_{21})\tau_{12}(I-T_{21}\tau_{12})^{-1}(\tau_{23}+T_{21}\tau_{13})T_0^{-1}+I \biggr)
%\end{eqnarray}
%\subsubsection{First Sub Subsection}

%\threesubsection{First lowest-level subsection}

\section{Conclusion}
In this work we construct the model of the regular array of the single-walled carbon nanotubes, which is simple enough and allows us to evaluate the phonon interference resulting to the Fano resonance in the presence of the locally resonance structures.
The latter can be formed by the additional nanotubes, which are placed over the CNT array in the various locations.
Varying the parameters of the nanotubes (the diameter, chirality and number of walls) we can change the position and the width of the destructive resonance, which results to the full reflection of the phonons with the certain frequency as in the acoustical as well as in the optical domain.
Also, in order to change the frequency interval we can modify the CNT's surface that leads to the changing the intertube constant $\chi$.
Thus, the model considered here can be useful in the investigations of the phonon as well as the electro-mechanical properties of the regular CNT structures.

%\medskip
%\textbf{Conflict of interest}

%The author declare no conflicts of interest.

% Acknowledgements
\medskip
%\textbf{Acknowledgements} \par %delete if not applicable))
%This work has ben performed under financial support by Russian Foundation for Basic Research (grant n. 19-58-45036 IND a)

% References
\medskip

\bibliographystyle{MSP}
\bibliography{CNT}

% Figures/tables and captions
% Permission statements are required for all figures reproduced or adapted from previously published articles/sources. Please also ensure that all necessary permissions to reproduce images have been received
% Please remove these statements for original figures

%
%\begin{table}
% \caption{Table 1 caption}
%  \begin{tabular}[htbp]{@{}lll@{}}
%    \hline
%    Description 1 & Description 2 & Description 3 \\
%    \hline
%    Row 1, Col 1  & Row 1, Col 2  & Row 1, Col 3  \\
%    Row 2, Col 1  & Row 2, Col 2  & Row 2, Col 3  \\
%    \hline
%  \end{tabular}
%\end{table}

% Please provide Biographies and photos for Essays, Feature Articles, Progress Reports, Reviews, and Perspectives for those authors who should be highlighted  
% These should be at most 100 words long
% For other article types this section can be removed
% Photographs should be 40mm broad and 50 mm high

%\begin{figure}
%  \includegraphics{bio-placeholder.jpg}
%  \caption*{Biography}
%\end{figure}
%
%\begin{figure}
%  \includegraphics{bio-placeholder.jpg}
%  \caption*{Biography}
%\end{figure}
%
%\begin{figure}
%  \includegraphics{bio-placeholder.jpg}
%  \caption*{Biography}
%\end{figure}
%
%\begin{figure}
%  \includegraphics{bio-placeholder.jpg}
%  \caption*{Biography}
%\end{figure}

% Table of contents entry should be 50 - 60 words long
% Image should be 55 mm broad and 50 mm high or 110 mm broad and 20 mm high

%\begin{figure}
%\textbf{Table of Contents}\\
%\medskip
%  \includegraphics{toc-image.png}
%  \medskip
%  \caption*{ToC Entry}
%\end{figure}

\end{document}